\documentclass[onecolumn]{aa}  

\usepackage{graphicx}
\usepackage{txfonts}
\begin{document}

   \title{Early Warning Signals Indicate a Critical Transition in Betelgeuse}
%

   \author{Sandip V George
          \inst{1}
          \and
         Sneha Kachhara
         \inst{2}
          \and
         Ranjeev Misra
         \inst{3}
          \and
          G. Ambika\inst{2}
          }

   \institute{University of Groningen, University Medical Center Groningen (UMCG), Groningen, Department of Psychiatry, Interdisciplinary Center Psychopathology and Emotion regulation (ICPE), The Netherlands .\\
              \email{s.v.george@rug.nl}
         \and
             Indian Institute of Science Education and Research (IISER) Tirupati, Tirupati - 517507, India\\
             \email{snehakachhara@students.iisertirupati.ac.in}\\
             \email{g.ambika@iisertirupati.ac.in}
             \and
             Inter University Centre for Astronomy and Astrophysics (IUCAA), Pune - 411007, India\\
             \email{rmisra@iucaa.in}
            }

   \date{}

 
  \abstract
   {Critical transitions occur in complex dynamical systems, when the system dynamics undergoes a regime shift. These can often occur with little change in the mean amplitude of system response prior to the actual time of transition. The recent dimming and brightening event in Betelgeuse occured as a sudden shift in the brightness and has been the subject of much debate. Internal changes or an external dust cloud have been suggested as reasons for this change in variability.  }
   {We examine whether the dimming and brightening event of 2019-20 could be due to a critical transition in the pulsation dynamics of Betelgeuse, by studying the characteristics of the light curve prior to transition.}
   {We calculate the quantifiers hypothesised to rise prior to a critical transition for the light curve of Betelgeuse up to the dimming event of 2019-20. These include the autocorrelation at lag-1, variance and the spectral coefficient calculated from detrended fluctation analysis (DFA), apart from two measures that quantify the recurrence properties of the light curve. Significant rises are confirmed using the Mann-Kendall trend test.}
   {We see a significant increase in all quantifiers (p < 0.05) prior to the dimming event of 2019-20. This suggests that the event was a critical transition related to the underlying nonlinear dynamics of the star. }
   {Together with results that suggests minimal change in $T_{eff}$ and infra-red flux, a critical transition in the pulsation dynamics could be a possible reason for the unprecedented dimming of Betelgeuse. The rise in the studied quantifiers prior to the dimming event, supports this possibility.}


   \maketitle
%

\section{Introduction}

   Betelgeuse or $\alpha$ Orionis is a fascinating supergiant whose variability has been of particular interest to astronomers. As a semi-regular variable star, Betelgeuse usually varies in visual magnitude from 0.6 to 1.1  approximately every 425 days, with some evidence of a longer 5.9 year period. (\cite{goldberg1984variability, samus2017general, dupree1987periodic,  stothers1971luminosities}). Being a rapidly evolving supergiant that is well off the main sequence makes this star particularly exciting to study. Quasi-hydrostatic evolutionary models predict that Betelgeuse will undergo a supernova explosion sometime in the next 100,000 years (\cite{dolan2016evolutionary}). 
   
   The interest in this star increased suddenly in the end of 2019 due to a large dimming and subsequent rapid brightening event that took place (\cite{guinan2020continued,sigismondi2020rapid}). The dimming reported by \cite{guinan2019fainting} re-ignited the questions related to the internal dynamics of this red supergiant (RSG) and led to speculations regarding an impending supernova explosion. Multiple hypothesis have been suggested in order to explain this dimming, which was recorded as the dimmest that the star has been in its observational history. An increase in the circumstellar dust
   and cooling in convective cells have been proposed as plausible explanations for the dimming phenomena (\cite{levesque2020betelgeuse}). Some authors have also noted that the dimming coincides with the minimum in the 2300 day and 400 day periodicities of the star (\cite{sigismondi2019betelgeuse,percy2020s}). The hypothesis that the dimming is due to variation in the convective cells has been refuted by a measurement of the surface temperature of the star, which showed no significant difference between the $T_{eff}$ measured in 2004 and near the minimum in 2020 (\cite{levesque2020betelgeuse}). They suggested that a dust cloud must be responsible for the rapid dimming of the star. The circumstellar envelope around Betelgeuse was previously observed and characterized by \cite{haubois2019inner}. \cite{gupta2020thermodynamics} further characterized the nature of the dust grains that could condense in this circumstellar envelope. While initial measurements by \cite{cotton2020multi} during the dimming phase indicated a reduction in polarisation, detailed analysis suggested that the polarised flux from the envelope remained constant during the dimming, and subsequently increased during the brightening (\cite{safonov2020differential}).  \cite{safonov2020differential} suggested that a dust cloud would result in an IR excess, close to the dimming. However, measurements by \cite{gehrz2020betelgeuse} during the dimming episode in the IR region, inferred that no significant change in flux was observed. This was further supported by observations in the sub millimeter wavelength by \cite{dharmawardena2020betelgeuse}.
   
   A critical transition occurs in a nonlinear dynamical system, when the nature of the system dynamics undergoes a drastic change. In dynamical systems theory, the point where this transition occurs is called a bifurcation point (\cite{hilborn2000chaos}). Very often, these critical transitions can take place with no visible variation in the mean amplitude in the system dynamics, prior to the point of actual transition, making these changes difficult to foresee. 
   
   One of the possibilities for the sudden dimming of Betelgeuse could be a critical transition in the pulsation dynamics of the star.  Pulsation has been long understood as a cause of variability in Betelgeuse (\cite{dupree1987periodic}). Variability due to stellar pulsations in many stars has been modelled using nonlinear dynamical models (\cite{baker1979pulsations,kollath2002nonlinear}). A transition in the pulsation dynamics of Betelgeuse can be captured using nonlinear time series analysis techniques on the light curve of the star. Nonlinear time series analysis has been used in multiple fields of astronomy, and stellar variability in particular (\cite{lindner2015strange, george2015effect, plachy2018chaotic}). An important application of nonlinear time series analysis has been to predict critical transitions in real world systems.

   A dynamical system can exist in a multitude of possible states, each of which may have a distinct pattern of evolution and exist for a particular set of system parameters. When the parameters change, the system dynamics transitions into another state.  Critical transitions in dynamical systems can take place even with minor changes in these parameter values. Such critical transitions would lead to major changes in the dynamical behavior of the system (\cite{hilborn2000chaos}). For instance, a Hopf bifurcation would lead to the birth of sudden periodic behavior. Seminal work by \cite{scheffer2009early} suggested that prior to a critical transition in a dynamical system, multiple time series quantifiers increase (\cite{scheffer2009early,scheffer2012anticipating}). In particular the autocorrelation at lag-1, the variance and the spectral exponent have been shown to increase prior to a transition. This has been used to predict dynamical transitions in many fields including ecology, engineering and psychiatry (\cite{dakos2012robustness,lenton2012early, trefois2015critical,ghanavati2014understanding,wichers2016critical, shalalfeh2016kendall}). \\
  Another way to detect a change in the dynamics of a system is to look at the associated phase space. Structures in phase space trace the evolution of the system in a way that can be quantified in terms of density and recurrence of points in time. The pattern of return or recurrences can be analyzed using what is known as Recurrence Quantification Analysis (RQA)(\cite{marwan2007recurrence}). This pattern of return times is known to change near critical transitions (\cite{wissel1984universal}), which reflects as changes in RQA measures.  RQA measures have been used for characterization of system dynamics in different domains such as climate studies (\cite{zhao2011identifying}), physiological data analysis (\cite{acharya2011application,marwan2002recurrence}), stock markets (\cite{bastos2011recurrence}) and engineering (\cite{godavarthi2017recurrence}). They have been particularly useful for the identification of dynamical transitions, including periodic-chaos, chaos-chaos transitions, as well as intermittent states (\cite{marwan2013recurrence,godavarthi2017recurrence}). 
  
  In stellar variability, critical transitions often take place at time scales that are much larger than the period of observation. The event of 2019-20 in Betelgeuse is a welcome exception. We examine the variation of these quantifiers in the light curve of Betelgeuse from 1990-2019 up to the the transition. We then check for significant variations in these quantifiers over time. If the dimming and subsequent brightening event of 2019-20 was due to internal factors leading to a critical transition in the pulsation dynamics of the star, we expect a detectable increase in these quantifiers. 
   
\section{Early warning signals in Betelgeuse}

Transitions in dynamics are characteristic of complex nonlinear dynamical systems, whereby the dynamics of the system undergoes a regime shift. Many of these sharp transitions are preceded by early warning signals (EWS) in the system response. A positive feedback loop drives the system forward to a bifurcation point, after a critical threshold is achieved (\cite{scheffer2009early}). Two quantifiers that are hypothesized to rise in this scenario are the variance and the autocorrelation at lag-1 (ACF(1)). A closely related quantifier is the detrended fluctuation analysis (DFA) exponent $\alpha$, which measures long term memory in the time series (\cite{peng1994mosaic}). Briefly, the detrended fluctuation analysis initially transforms the data into a time series of the cumulative amplitude distribution. The fluctuation (determined as the root mean squared deviation) of this cumulative time series from the linear trend is calculated at different time scales. This  fluctuation is expected to rise as a power-law with respect to the timescale considered, the exponent of which is the Hurst exponent, $\alpha$ (\cite{peng1994mosaic, livina2007modified, shalalfeh2016kendall}). It may be worthwhile to notice that all transitions are not preceded by early warning signals. Trivially, transitions caused by external influences would not be preceded by any kind of warning signal from the system dynamics. Moreover, not all dynamical transitions are preceded by  warning signals either. Hence early warning signals are broad indicators of a possible upcoming transition (\cite{scheffer2012anticipating}).
   
For our analysis we use light curve data from the American Association of Variable Star Observers (AAVSO) (\cite{Kafka2020}). The light curve of Betelgeuse is binned in ten day bins to examine the long trends in the data. As mentioned earlier we quantify the lag-1 autocorrelation, the variance and the Hurst exponent $\alpha$ (\cite{scheffer2009early,livina2007modified}). 
The quantifiers are calculated over moving windows of size 300 points. The window is moved forward by 1 data point, irrespective of gaps. A study across different window sizes is conducted, and the results are shown in the appendix \ref{app:window}. While the effect sizes reduced for individual quantifiers, especially at low window sizes, the broad conclusions remain unchanged.
All calculations were conducted in python 3.5.2 using the numpy, scipy and entropy packages (\cite{oliphant2006guide, virtanen2020scipy,Vallat2020}).

Missing data is a cause for concern in this dataset, as Betelgeuse is not observable for a large period between May and July. However, if stationarity is assumed at the time scales of the size of the gaps, calculations of the variance and ACF(1) should not be affected much by missing data. Further, the Hurst exponent $\alpha$, has been shown to be very robust to missings in positively correlated data (\cite{ma2010effect}). The profile of gaps in the AAVSO dataset, and its variation with time is considered in appendix \ref{app:gaps}. Due to the  high prevalence of gaps in the period from 1980-1990, trends are analyzed for the period starting from 1990 onwards.

\begin{figure}
    \centering
    \includegraphics[width=.45\textwidth]{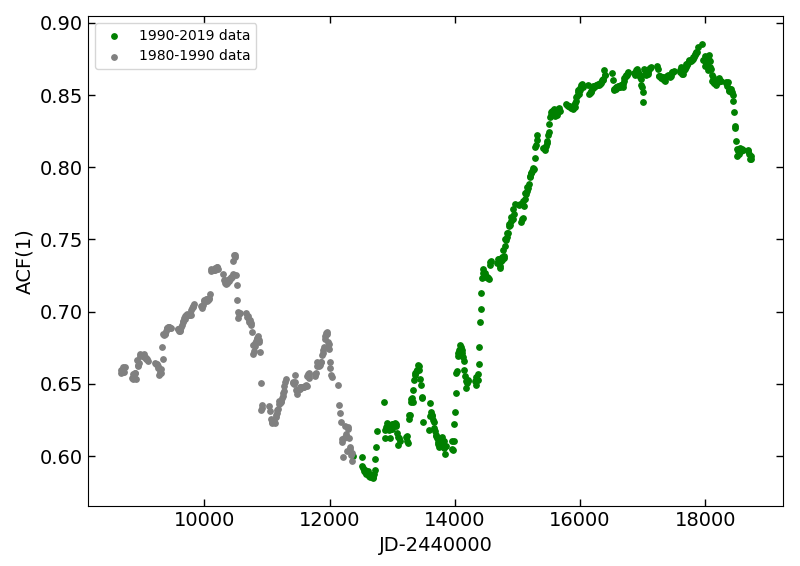}
    \includegraphics[width=.45\textwidth]{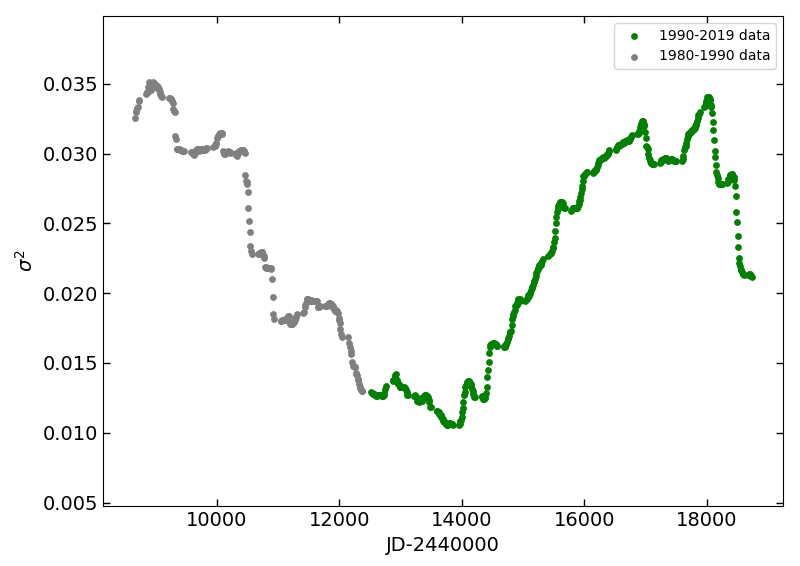}
    \includegraphics[width=.45\textwidth]{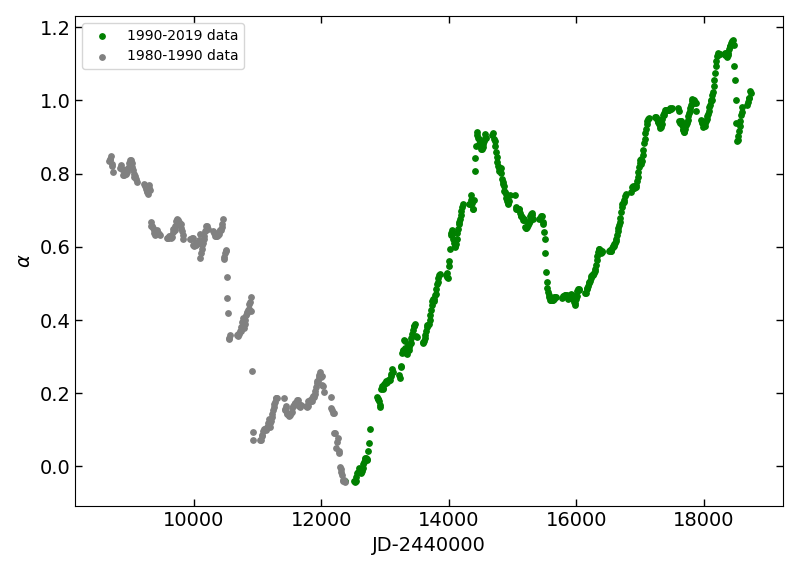}
    \caption{Variation of the autocorrelation at lag-1, variance and $\alpha$ calculated from detrended fluctuation analysis, prior to the dimming event. The period from 1980-1990 is shown in grey and the period from 1990 onward is shown in green. A visible rise can be seen leading towards the event. The calculated quantifier is plotted at the end time of the moving window. }
    \label{fig:acf_1}
\end{figure}

   To determine trends in the quantifiers, we examine the Kendall correlation coefficient ($\tau_K$) up to the transition point. To correct for correlation due to the moving window approach, we use the Hamed-Rao correction (\cite{hamed1998modified}). Calculations were conducted using the pymannkendall package in python (\cite{Hussain2019pyMannKendall}). 
   Increasing trends are seen for all three quantifiers with time, prior to the transition in brightness, indicating that a dynamical transition in Betelgeuse led to the 2019-20 dimming and brightening event. The trends, values of $\tau_K$ and p-values are listed in Table \ref{table:ews}. The change in the EWS with time are shown in Figure \ref{fig:acf_1}. The grey region (1980-90 data) in Figure \ref{fig:acf_1} shows an increased variance and $\alpha$ prior to the period considered. \cite{joyce2020standing} point out a major dimming event in the mid to late 1980s, which may be a possible explanation for the change in these quantifiers around that period. 
   
   In the next section, we will supplement the warning signals detected in this section with quantifiers derived using recurrence quantification analysis.

   \begin{table}
\caption{Significant increases in early warning signals prior to the dimming event in Betelgeuse data. Kendall-$\tau$ correlation coefficient ($\tau_K$), p-value and significance calculated using the modified Mann-Kendall test applied after corrections \cite{hamed1998modified}.}             
\label{table:ews}      
\centering                          
\begin{tabular}{c c c c}        
\hline\hline                 
Quantifier  &Trend & $\tau_{K}$  & $p-value$  \\ [0.5ex] 
\hline 
$ACF(1)$ & increase & $0.752$ & $<0.001$  \\ 
$Variance$  & increase & $0.653$ & $<0.001$ \\ 
$\alpha$  & increase & $0.505$ & $0.008$  \\ 
\hline                                   
\end{tabular}
\end{table}
%

\section{Recurrence Based analysis}
   
   Recurrence based measures have been applied for the study of black holes (\cite{jacob2018recurrence}), variable stars (\cite{george2019classification}), solar radiation (\cite{ogunjo2017investigating}) and exoplanetary systems (\cite{kovacs2019recurrence}. A rather remarkable advantage of RQA measures is their ability to detect dynamical transitions from a time series even when the details of the underlying dynamics are elusive (\cite{marwan2013recurrence,thiel2004estimation}). These studies show that recurrence based measures are successful even with non-stationary and short datasets. \\
   The time series is first transformed to an attractor in phase space, by what is known as delay embedding (\cite{kantz2004nonlinear,takens1981detecting}). The process generates an attractor in phase space which is topologically equivalent to the original attractor and captures evolution of the system for the duration of given time series. Every point on the attractor is considered a vector and a distance matrix is constructed which encapsulates recurrences in the system in terms of distances between the vectors. If we apply a threshold to the distance metric such that the points within the threshold are considered close, then we get the recurrence matrix, $R$. Each element of $R$ corresponds to a unique pair of points in the phase space, and the value of 1 indicates their proximity (0 otherwise). The visual representation of $R$, with 0s as white dots and 1s as black dots is a Recurrence Plot (RP).

  \begin{figure}
    \centering
    \includegraphics[width=.45\textwidth]{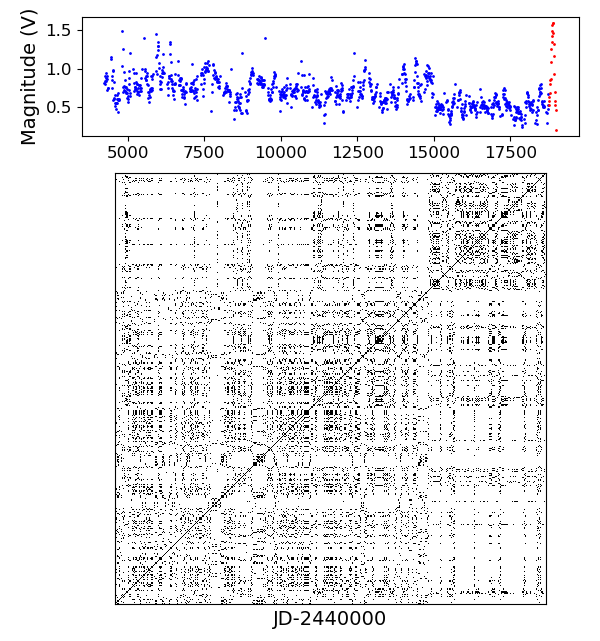}
    \caption{Recurrence Plot of the time series of Betelgeuse. The upper panel shows corresponding time series (in blue). The red part (dimming) was not included in the analysis. The parameters used are $m$ = 1, $\tau$ = 1, fixed RR = 0.1}
    \label{fig:RP}
   \end{figure}
   
   The RP for the entire time series of Betelgeuse for the duration 1980 up to just before the dimming event in 2019 is shown in the figure \ref{fig:RP} for illustration. The moving window approach considers segments of the time series, however, and the corresponding RPs are constructed for each window. We fix the recurrence threshold such that the recurrence rate stays at 0.1 and calculate other quantifiers. The exact definitions along with the parameters used for the construction of RP can be found in the appendix \ref{app:rp}. \\
   
   RQA measures assess the patterns formed by points, diagonal lines, vertical lines, and other structures in the RP. RQA measures such as Recurrence Rate (RR), Determinism (DET), Laminarity (LAM) etc. are intimately related to the underlying dynamics: RR computes number of recurred states (based on the number of recurrence points), DET estimates how deterministic the dynamics appears to be (based on the distribution of diagonal lines in RP) and LAM reflects the extent of laminar phases in the system (based on distribution of white vertical lines), or intermittency. 
   In the case of a dynamical transition, the values should deviate significantly from an overall average calculated with the assumption of no transition (\cite{marwan2013recurrence,schinkel2009confidence}).\\
   
   \begin{figure}
    \centering
    \includegraphics[width=.45\textwidth]{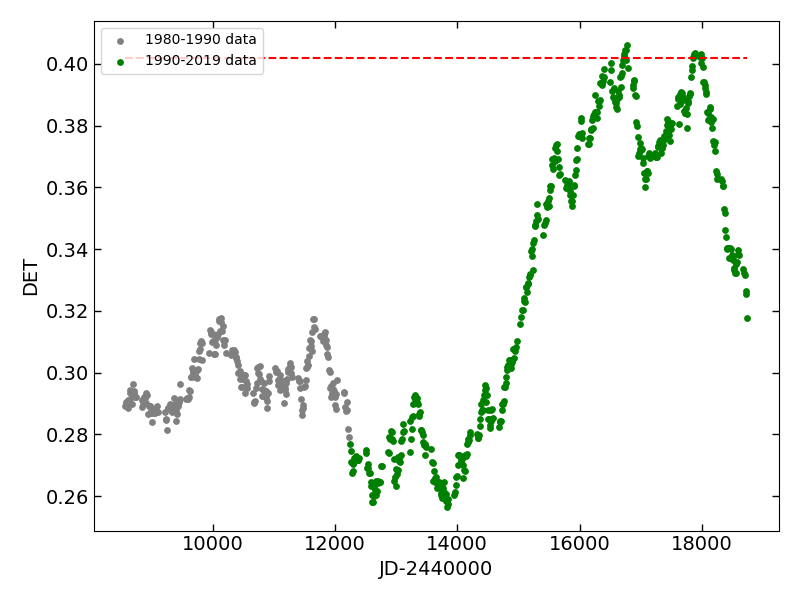}
    \includegraphics[width=.45\textwidth]{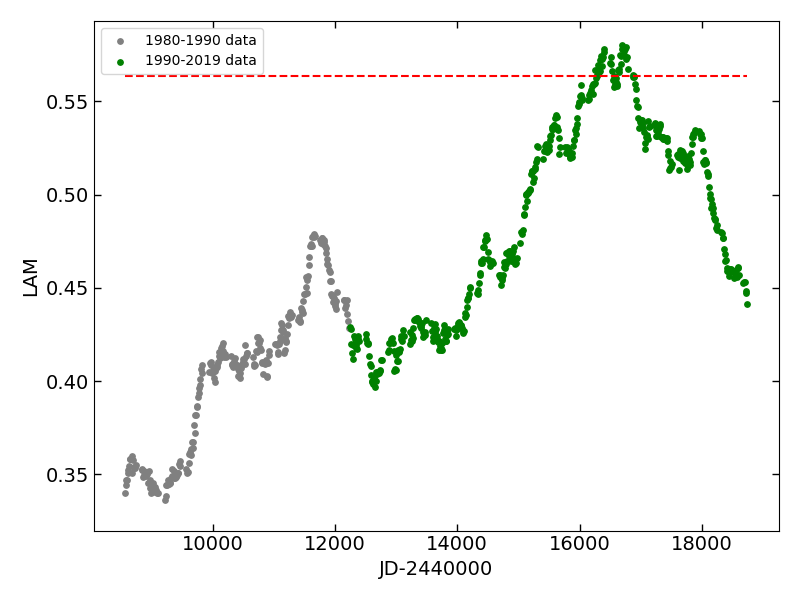}
    \caption{Variation of the RQA measures DET and LAM, prior to the dimming event for fixed RR = 0.1. The red line indicates 95\% confidence level. The color code is similar to figure \ref{fig:acf_1}, with grey region referring to 1980-1990.} 
    \label{fig:RQA}
\end{figure}
   
   \begin{table}
\caption{Significant increases in RQA measures prior to the dimming event in Betelgeuse data. Kendall-$\tau$ correlation coefficient ($\tau_K$), p-value and significance calculated using the modified Mann-Kendall test applied after corrections \cite{hamed1998modified}.}             
\label{table:ews_rqa}      
\centering                          
\begin{tabular}{c c c c}        
\hline\hline                 
Quantifier  &Trend & $\tau_{K}$  & $p-value$  \\ [0.5ex] 
\hline 
$DET$ & increase & $0.605$ & $<0.001$  \\ 
$LAM$  & increase & $0.457$ & $0.028$ \\ 

\hline                                   
\end{tabular}
\end{table}
   The behaviour of these measures is shown in figure \ref{fig:RQA}. We notice that both LAM and DET increase systematically as we approach the dimming episode and appear to follow overall a similar trend. The results quantifying the change in the DET and LAM measures with time using the modified Mann Kendall test is shown in Table \ref{table:ews_rqa} (\cite{hamed1998modified}). The test confirms a significant gradual increase in the measures. We observe that both measures DET and LAM go beyond the 95\% confidence level (obtained from 1000 bootstrappings as suggested by \cite{schinkel2009confidence}) before the dimming with LAM rising slightly before DET. Such a behaviour can be expected from intermittency (reflected in the high value of LAM) associated with the phenomenon of critical slowing down leading to a dynamical shift (significant change in DET).

\section{Results and Discussion}

In this work we analyze the light curve of Betelgeuse from the 1980s up till prior to the dimming event. Our analysis suggests that signatures of an impending change in the nonlinear dynamics can be observed from the properties of the light curve preceding the dimming episode. We use quantifiers that are used in early warning signal literature, supplemented with recurrence based analysis to show conclusively that certain properties of the light curve showed a significant change leading to the dimming event in 2019.

The rise in early warning signals prior to the dimming episode in late 2019, has been quantified using the Mann-Kendall test, which checks the correlation of the quantifiers with time. Along with classically used early warning quantifiers, we also check the variation of the recurrence based quantifiers. These independently show a change prior to the dimming event. A comparison between the two suggests that the change in quantifiers start at about the same time.

This result has significant impact on our understanding of the 2019 dimming event. The observations in the sub-millimeter wavelengths reported by \cite{dharmawardena2020betelgeuse} and in the IR by \cite{gehrz2020betelgeuse} suggest that the dimming is not compatible with a dust cloud. Our results add to this conclusion by suggesting that the dimming event is best explained by a change in the intrinsic dynamics of the star.
\cite{levesque2020betelgeuse} showed that there was no significant decrease in $T_{eff}$ during the dimming episode, which largely eliminates a convection driven dimming. The final option then suggests a dimming episode that is driven by a change in the pulsation dynamics. A pulsation driven dimming episode of about one magnitude reduction would not cause a significant reduction in $T_{eff}$ and has already been proposed a probable cause by \cite{dharmawardena2020betelgeuse}. Since much of the flux in Betelgeuse is emitted in the IR region, small changes in $T_{eff}$ can lead to large changes in emitted visible light spectrum (\cite{karttunen2016fundamental, reid2002mira}). Hence dimming due to a critical transition in the stellar pulsation dynamics can happen with minimal change in the IR spectrum, consistent with observations on Betelgeuse during the dimming  (\cite{gehrz2020betelgeuse,reid2002mira}). 
   
   \section{Conclusions}
The reasons behind the dimming and brightening event in Betelgeuse during 2019-20 still remain largely unclear. Stochastic chance variations, oscillatory phenomena etc are various possibilities that could explain this event. A critical transition in the pulsation dynamics of the star is another possibility. The observation of early warning signals in the Betelgeuse light curve well before the dimming event suggests that it could be caused by the latter. An increase in early warning signals is thought to be suggestive of an impending critical transition in the system (\cite{scheffer2009early}). Our analysis shows significant increases in the traditional early warning signal quantifiers as well as in recurrence plot based quantifiers, prior to the dimming event. As opposed to a transient dimming episode, a critical transition would imply a permanent change in the behavior of the system. As Betelgeuse becomes visible again, more data may throw fresh light on the mystery that surrounds this event.

\begin{acknowledgements}
      We acknowledge with thanks the variable star observations from the AAVSO International Database contributed by observers worldwide and used in this research. 
      
      SVG acknowledges financial support from the European Research Council (ERC) under the European Union’s Horizon 2020 research and innovative programme (ERC-CoG-2015; No 681466 awarded to M. Wichers).
      
      SK acknowledges financial support from the Council of Scientific and Industrial Research (CSIR), India. 
      Part of the analysis was performed with the help of the pyunicorn package (\cite{donges2015unified}) available at http://www.pik-potsdam.de/~donges/pyunicorn/.
\end{acknowledgements}

%
%

\begin{appendix}

\section{Datagaps}
\label{app:gaps}
In this appendix we will briefly describe the analysis on the prevalence of datagaps in the time series. The lack of visibility of Orion from May to July results in periodic gaps in the light curve. Work by \cite{george2015effect} suggests that two distributions are important to fully describe the prevalence of datagaps in a time-series, namely the distributions of the gap size and gap frequency. The mean gap size for the entire time series is 19.14 days and the mean gap position (the average time between two gaps) is 20.98 days. 

\begin{figure}
    \centering
    \includegraphics[width=.35\textwidth]{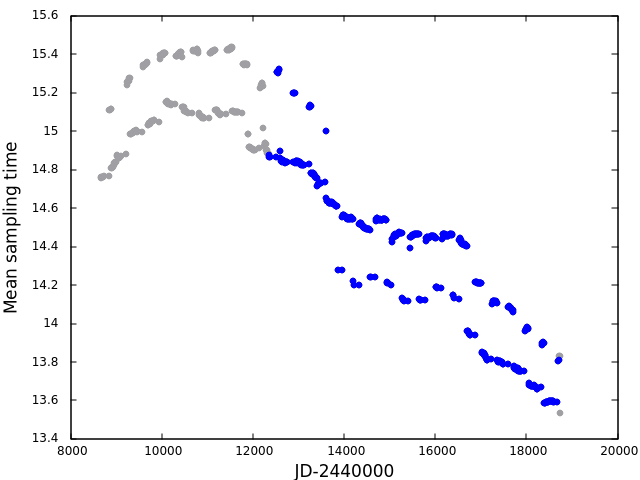}
    \includegraphics[width=.35\textwidth]{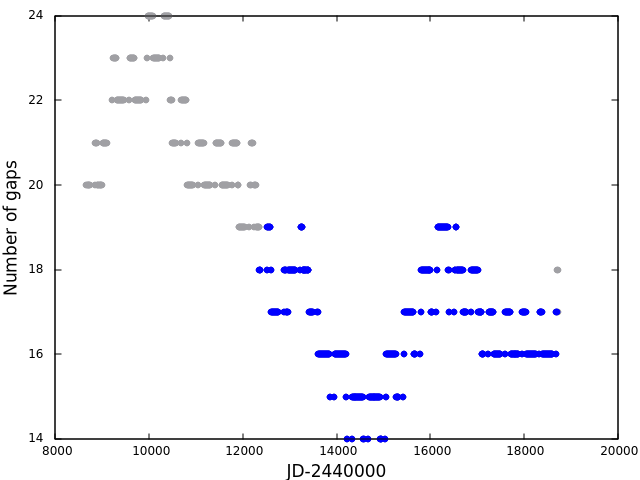}
    \caption{Variation of (a) mean sampling time and (b) number of gaps of the light curve over time. The grey dots represent the period from 1980-1990 and the blue shows the period from 1990 onwards. The oscillation in (a) is due to the periodic gaps present due to lack of visibility of $\alpha$-Orionis during part of the year.}
    \label{fig:gaps}
\end{figure}

For our analysis it is important to check whether the gap distributions are stationary. Hence we examine the trends in two quantities. The average sampling time in windows and the average number of gaps in windows. These are shown in Figure \ref{fig:gaps}. We find a higher prevalence of gaps in earlier periods in the AAVSO data. Hence our analysis is conducted from the period starting 1990. 

\section{Construction of Recurrence Plots}
\label{app:rp}
In this appendix we briefly describe the construction process of Recurrence Plots (RPs) and discuss the parameters used. The process consists of two steps: embedding and calculation of recurrence matrix. 
The process of embedding leads to a reconstructed attractor (collection of trajectories) in the phase space where each point, or vector $v_{i}$ represents a microstate. From the given time series $y_{t}$, vectors $\vec{v}_{i}$ are constructed as follows:

\[\vec{v}_{i} = [y(t_{i}),y(t_{i}+\tau),y(t_{i}+2\tau),..,y(t_{i}+(m-1)\tau)]\]

where $m$ is the embedding dimension(dimension of the phase space) and $\tau$ is the delay (\cite{takens1981detecting,kantz2004nonlinear}). Studies show that properties of RPs don't vary significantly with embedding parameters (\cite{thiel2004estimation}). We chose $m$ = 1, $\tau$ = 1. 

Once we have the vectors $\vec{v}_{i}$, the next step is to construct the recurrence matrix $R$ as follows:

\[R_{ij} = \Theta \left ( \varepsilon-\left \| \vec{v_{i}}- \vec{v_{j}} \right \| \right )\]

where $\Theta$ is the Heaviside step function. $||...||$ represent a distance norm (supremum norm used here) between vectors $\vec{v}_{i}$ and $\vec{v}_{j}$, and $\varepsilon$ is the distance threshold. The RP represents $R$ visually, with 0s as white spaces and 1s as black. 
The Recurrence Rate (RR) is given as:

\[RR = \frac{1}{N^{2}}\sum_{i,j=1}^{N}R_{ij}\]
where N is the total number of points. We adopt variable $\varepsilon$ value for a fixed $RR$ = 0.1. 
The RQA measures are then calculated using $R$.

DET captures the structure and extent of diagonal lines, indicating the degree to which one can 'predict' the system:

\[DET = \frac{\sum_{l=l_{min}}^{N}lP(l))}{\sum_{l=1}^{N}lP(l)}\]

where $l$ is the length of a diagonal line and $P(l)$ is the histogram of diagonal lines.

LAM captures the structure and extent of vertical lines, reflecting the laminar phases in the system. A laminar phase in the system represents a micro-regime where the system spends a lot of time. It is given as:

\[LAM = \frac{\sum_{v=v_{min}}^{N}vP(v))}{\sum_{v=1}^{N}vP(v)}\]

where $v$ is the length of a vertical line and $P(v)$ is the histogram of vertical lines.

\section{Variation with window size}
\label{app:window} 
In this appendix, we consider variation of standard early warning signals with changing window size. We vary the window sizes for bin size 10 from 100 to 500 for the dataset from 1990 onwards. The results are presented in Tables \ref{table:window} and \ref{table:window_rqa}, for standard early waning signal measures and recurrence quantification measures respectively. The Kendall-$\tau$ coefficient are calculated using the the scipy package, while the corrected significance values are calculated using the pymannkendall package (\cite{virtanen2020scipy,Hussain2019pyMannKendall}). Broadly one sees that the conclusions drawn in the main text of this paper holds for a range of window sizes. 
   \begin{table*}
\caption{ Kendall-$\tau$ correlation coefficient ($\tau_K$), p-value before and after applying the corrected Mann-Kendall test (\cite{hamed1998modified}) for the ACF, Variance and $\alpha$ from the detrended fluctation analysis.}             
\label{table:window}      
\centering                          
\begin{tabular}{c c c c c c c c}        
\hline\hline                 
Window Size  & ACF $\tau_{K}$  & $p-value$ (corrected) & Variance $\tau_{K}$  & $p-value$ (corrected) & $\alpha$ $\tau_{K}$  & $p-value$ (corrected)  \\ [0.5ex] 
\hline 
$100$ & $0.206$ & $<0.001 (.192)$  & $-0.004$ & $ 0.874 (0.978)$ & $0.105$ & $<.001 (0.412)$ \\
$200$ & $0.394$ & $<0.001 (.055)$  & $0.224$ & $ <.001 (0.224)$ & $0.318$ & $<.001 (0.094)$ \\
$300$ & $0.752$ & $<0.001 (<.001)$  & $0.653$ & $<0.001 (<.001)$ & $0.505$ & $<0.001 (.008)$  \\
$400$ & $0.892$ & $<0.001 (<.001)$  & $0.889$ & $<0.001 (<.001)$ & $0.597$ & $<0.001 (<.001)$  \\
$500$ & $0.769$ & $<0.001 (<.001)$  & $0.784$ & $<0.001 (<.001)$ & $0.061$ & $0.146 (.557)$  \\
\hline                                   
\end{tabular}
\end{table*}
 \begin{table*}
\caption{ Kendall-$\tau$ correlation coefficient ($\tau_K$), p-value before and after applying the corrected Mann-Kendall test (\cite{hamed1998modified}) for the determinism (DET) and laminarity (LAM) measures from recurrence quantification analysis.}             
\label{table:window_rqa}      
\centering                          
\begin{tabular}{c c c c c c}        
\hline\hline                 
Window Size  & DET $\tau_{K}$  & $p-value$ (corrected) & LAM $\tau_{K}$  & $p-value$ (corrected)  \\ [0.5ex] 
\hline 
$100$ & $0.004$& $0.884 (0.939)$  & $0.122$ & $ <0.001 (0.461)$ \\
$200$ & $0.336$ & $<0.001 (0.062)$  & $0.188$ & $ <0.001 (0.356)$  \\
$300$ & $0.613$ & $<0.001 (<0.001)$  & $0.465$ & $<0.001 (<.028)$  \\
$400$ & $0.787$ & $<0.001 (<0.001)$  & $0.657$ & $<0.001 (<.001)$ \\
$500$ & $0.864$ & $<0.001 (<0.001)$  & $0.773$ & $<0.001 (<.001)$  \\
\hline                                   
\end{tabular}
\end{table*}

\end{appendix}

\end{document}